\documentstyle[epsfig]{elsart}
\newcommand{\bea}{\begin{eqnarray}}
\newcommand{\beq}{\begin{equation}}
\newcommand{\eea}{\end{eqnarray}}
\newcommand{\eeq}{\end{equation}}

\begin{document}
\runauthor{L. Benet et al.}

\begin{frontmatter}
\title{Semiclassical properties of eigenfunctions and occupation
number distribution for a model of two interacting particles}

\author[ccf,cic]{L. Benet\thanksref{LBF}},
\author[cic,buap]{F.M. Izrailev},
\author[ccf,cic]{T.H. Seligman} and
\author[if]{A. Su\'arez-Moreno}

\address[ccf]
{Centro de Ciencias F\'{\i}sicas, U.N.A.M., Cuernavaca, M\'exico}
\address[cic]
{Centro Internacional de Ciencias, Cuernavaca, M\'exico}
\address[buap]
{Instituto de F\'{\i}sica, B.U.A.P., Apdo. Postal J-48,
72570 Puebla, M\'exico}
\address[if]
{Instituto de F\'{\i}sica, U.N.A.M., M\'exico D.F., M\'exico}

\thanks[LBF]{Present address: Max-Planck-Institut f\"ur Kernphysik,
D-69117 Heidelberg, Germany. Email: Luis.Benet@mpi-hd.mpg.de}

\date{\today}

\begin{abstract}
Quantum-classical correspondence for the shape of eigenfunctions,
local spectral density of states and occupation number
distribution is studied in a chaotic model of two coupled quartic
oscillators. In particular, it is shown that both classical
quantities and quantum spectra determine global properties of
occupation numbers and inverse participation ratio.

\end{abstract}

\end{frontmatter}


Recently, the study of the quantum manifestations of classical
chaotic systems (quantum chaos) has turned from spectral
statistics to properties of eigenfunctions. For the former,
statistical aspects of spectral fluctuations are well established:
Random matrix predictions follow for chaotic systems and
Poisson-like statistics for the integrable
ones~\cite{eigenvalues,rmt}. For eigenfunctions, however, the approach
seems not so straightforward. The inherent difficulty here arises
essentially from the dependence on the basis. This forces us
either to define basis independent quantities, or to specify a
basis.

In this letter we shall follow the second possibility, trying to
develop a framework as general as possible. We shall concentrate
on the quantum-classical correspondence of the quantities
such as the shape of eigenfunctions (EF), the local spectral
density of states (LDOS) and the single-particle occupation number
distribution ($n_s$). We study a Hamiltonian that displays classical 
chaos and has a spreading width in the single particle basis, that 
is sufficiently large to allow statistical treatment of the components.
We find excellent agreement between these quantities and their 
classical analogues in the semiclassical region. The classical 
analogues are defined through phase space integrals. Therefore, they 
do not depend on dynamical properties such as integrability or chaos 
of the full Hamiltonian. We shall also show that this correspondence 
allows one to approximately obtain some important characteristics 
for which quantum phases of eigenfunctions play no role. In
this way, in the semiclassical limit, one can obtain mean values
of single-particle operators without diagonalization of large
matrices. In particular, we present calculations for the inverse
participation ratio which distinguishes localized states from
extended ones in the unperturbed basis. These computations require
only the knowledge of the classical analogue of EF and the spectra
of the unperturbed Hamiltonian. We present our approach for the
two-body problem, but essential parts of the approach can be
easily extended to the $N$-body problem.

Let us begin with considering a two-body Hamiltonian of the form
$H=H_0+V$, and assume that its classical dynamics is fully
chaotic. Here, the unperturbed Hamiltonian is separable and,
therefore, integrable in terms of two one-particle Hamiltonians,
i.e. $H_0=h_1+h_2$. In turn, $V$ is the potential which couples
the motion of the particles. For our purposes, we shall assume
that both $H$ and $H_0$ remain invariant under the particle
interchange ($h_1=h_2=h$).

For the quantum treatment of such Hamiltonians the unperturbed
basis $H_0$ seems a convenient choice for the representation of
exact eigenfunctions. The rate of convergence certainly depends on
the strength of the perturbation $V$. For instance, if exact
eigenfunctions are extended all over the energy range considered,
as it is the case for certain potentials near the dissociation
limit~\cite{indues}, the convergence will be very slow. We,
therefore, assume that $V$ is such that the perturbed
eigenfunctions are extended over a certain energy range as to
allow convergence, but that this range is large enough in terms of
the number of principal components, i.e. that the spreading width
is sufficiently large~\cite{BSW93}.

The unperturbed basis is defined by the eigenfunctions
$|\Phi^0_k\rangle$ of $H_0$, reordered according to increasing 
eigenvalues, $E^0_k<E^0_l$ for $k<l$. This basis functions are 
written as properly symmetrized linear combinations of products of
single-particle basis states. The single-particle basis is defined
by the Schr\"odinger equation, 
$h|\varphi_i\rangle=\epsilon_i|\varphi_i\rangle$, and the combination 
of the basis states is such that $E^0_k=\epsilon_{i_1}+\epsilon_{i_2}$. 
In this sense, the basis defined by the mean field approximation 
coincides with the unperturbed basis when $V$ is the residual 
interaction.

Denoting by $|\Psi_i\rangle$ the eigenstates of the total
Hamiltonian and by $E_i$ the corresponding eigenvalues, in terms
of the basis states we have
\beq
\label{expansion}
|\Psi_i\rangle=\sum_k C^i_k |\Phi^0_k\rangle.
\eeq
The expansion coefficients $C^i_k$ define some global quantities
that we consider here. First, the shape of eigenfunctions (EF),
also called the F-function, is defined as the distribution
obtained by an average of the squared expansion coefficients as a
function of the unperturbed energy~\cite{izr1}
\beq
F^{i}_k \equiv \overline{{\left |C^{i}_k\right |}^2}
    =F(E_i,E^0_k).
\label{swidth}
\eeq
Here, the average is defined over a small window of perturbed
eigenstates around $E_i$. 
The average has been introduced in order to smooth the fluctuations
arising from individual wave functions considered.
It has been shown that the
F-function defines a kin of thermodynamic partition function for
systems of finite number of interacting particles~\cite{izr1}, if 
the components meet certain statistical requirements.
These will certainly be met, if the classical Hamiltonian leads to 
chaotic motion and the spreading width is large enough.

The second quantity of our interest, the LDOS gives the distribution
of unperturbed eigenstates in terms of the perturbed ones. The
LDOS is related to the EF by~\cite{izr1}
\beq
\label{ldos}
P_k^{i} \equiv F(E_i,E^0_k) \rho(E_i),
\eeq
where $\rho(E_i)$ is the level density for exact eigenstates, and
the F-function is taken now for a fixed value of the unperturbed
energy $E^0_k$. Therefore, the LDOS is a function of the perturbed
energy $E_i$.

These quantities have well-defined classical interpretations. For
instance, the classical EF is the distribution resulting from the
time-dependent unperturbed energy ${\cal E}_0(t)$, obtained by
substituting the solutions of the equations of motion for the
Hamiltonian $H$ into the expression for $H_0$. Since $H$ is
assumed to generate fully chaotic and thus ergodic dynamics, we
can replace the time integration along one typical orbit by a phase
space integral. Therefore, one can write for the classical EF the
phase space integral
\beq
\label{classical}
g({\cal E}, {\cal E}_0)=
A \int {\rm d}{\bf p}{\rm d}{\bf q}
\delta({\cal E}-H({\bf p},{\bf q}))
\delta({\cal E}_0-H_0({\bf p},{\bf q})),
\eeq
where ${\bf q}=(q_1, q_2)$, ${\bf p}=(p_1, p_2)$ are the position
and momentum vectors, and $A$ is a normalization constant.
For the classical EF in Eq.~(\ref{classical}), the independent
variable is ${\cal E}_0$; the total energy ${\cal E}$ is fixed.

The classical LDOS can be obtained in the same terms by
integrating the equations of motion for $H_0$ and substituting the
solutions into the expression for $H$. Since $H_0$ is an
integrable Hamiltonian, one is forced to consider an average over
different initial conditions. Then, Eq.~(\ref{classical}) serves
also to define the classical LDOS; ${\cal E}_0$ is now held fixed
and ${\cal E}$ is the independent variable. In the following, we
shall use the notation $g({\cal E}_0)$ to indicate the
classical EF and $g({\cal E})$  the classical LDOS, when
referring to Eq.~(\ref{classical}). In this notation, the variable
that is explicitly written is the independent variable.

We notice that the classical and quantum quantities, for instance
the EF as given by Eq.~(\ref{classical}) and the F-function
Eq.~(\ref{swidth}), differ in the way they are normalized.
The former, being a probability distribution, is normalized according
to $\int g({\cal E}, {\cal E}_0) d{\cal E}_0 =1$, which actually
defines the value of the constant $A$.
For the latter, the normalization is the unitarity condition for the
expansion coefficients, i.e. $\sum_k |C_k^i|^2=1$.
For a fair comparison of the classical and quantum results we require
the normalizations to be of the same type.
This is achieved including the local density of states, i.e.
$\sum_k\int |C_k^i|^2 \delta(E^0-E^0_k) dE^0=1$.
Numerically, we shall calculate this expression replacing the local
density of states by a step function, which is different from zero
only in a small interval that includes some levels, where its value
is one.
Then, we divide the energy range in a number of bins, and associate
to each bin the sum of the intensities $|C_k^i|^2$ of the energy
levels contained in it.

We shall turn now to the comparison among the quantum EF and LDOS with
their classical counterparts in the following model.
We consider two indistinguishable coupled quartic oscillators with
the Hamiltonian
\beq
\label{model}
H  = {1\over 2}(p_1^2+p_2^2)+ \alpha(x_1^4+x_2^4)+
\beta x_1^2x_2^2+\gamma (x_1^3x_2+x_1x_2^3).
\eeq
Here, $\alpha>0$ and we consider $\beta<0$ in order to have strongly
chaotic dynamics far from the dissociation limit, which is given
by $2\alpha+\beta+2|\gamma|=0$. For the results presented below,
we have used $\alpha=10$, $\beta=-5.5$ and $\gamma=5.6$; the
system is strongly chaotic and the phase space is quite
homogeneous~\cite{preparation}.

The system defined by the Hamiltonian~(\ref{model}) is obviously
integrable for $\beta=\gamma=0$; we shall consider this case to
define the unperturbed Hamiltonian $H_0$. We note that in the
basis defined by $H_0$ there are diagonal contributions from the
term $\beta x_1^2x_2^2$. These contributions could be incorporated
in the definition of $H_0$ in order to improve the approximate
mean field, but we avoid this complication.

Since the potential is a homogeneous polynomial, the system scales
classically with the energy.
This property can be carried over to the classical expressions
for the EF, LDOS and $n_s$.
For instance, for the classical EF, one finds
$g({\cal E},{\cal E}_0)={\cal E}^{-1} g(1,{\cal E}_0 / {\cal E})$,
with obvious extensions to the other quantities.

\begin{figure}
\noindent\centerline{
\psfig{figure=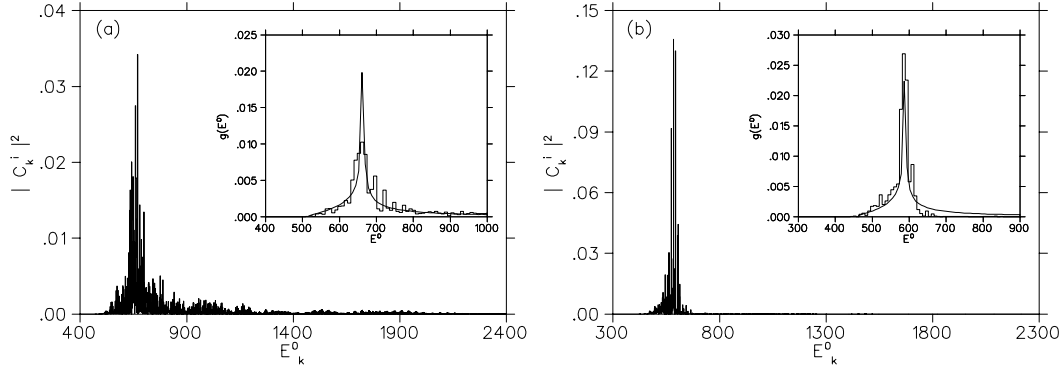,height=\textwidth,angle=90}}
\caption{
\label{fig1}
EF for (a)~a typical eigenstate $E_{620}=664.073$ and, (b)~for
a localized one $E_{515}=588.540$.
Notice the different vertical scales.
Insets: comparison between the classical (continuous curve) and
the quantum results (histogram).}
\vskip0.5cm
\end{figure}

In the following we present results obtained for fermions of even
parity for the Hamiltonian~(\ref{model}); similar results were
obtained for other symmetry classes. Figure~\ref{fig1}a refers to
the EF of a typical eigenstate, which extends over a certain range
of unperturbed energies; Fig.~\ref{fig1}b shows the EF for the
uncommon case of an eigenfunction with a smaller number of principal
components, a localized eigenfunction. This can be readily
appreciated in the vertical scale (intensity) and on the apparent
density of peaks (the eigenfunctions are normalized). The
distinction between a localized and an extended eigenstate can be
made quantitative, for instance, by considering the number of
principal components $l$: In the former case we obtain $l=(\sum_{k}
|C_k^{i}|^4)^{-1}=146.5$, while for the latter we have $l=18.4$.
Similar properties are also found when we
consider the LDOS for individual eigenfunctions of $H_0$.

\begin{figure}
\noindent\centerline{
\psfig{figure=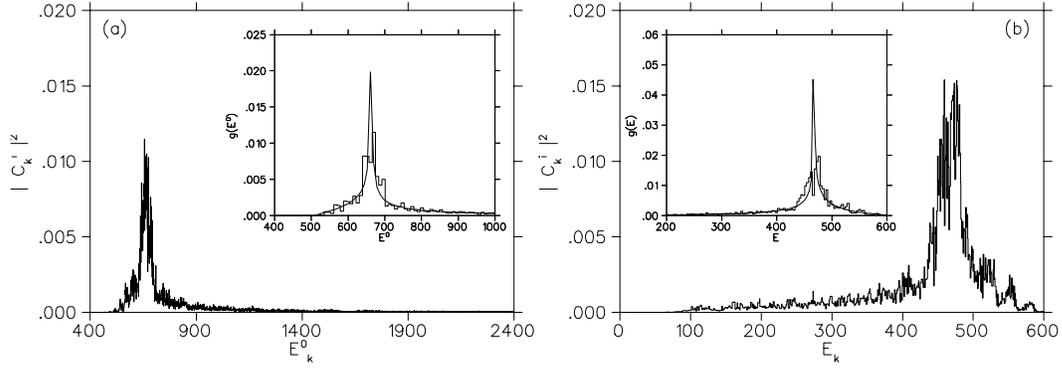,height=\textwidth,angle=90}}
\caption{
\label{fig2}
Results for a recentered average involving 21 neighbouring
eigenstates: (a)~EF around $E^0_{620}$; (b)~LDOS around
$E^0_{300}=469.102$. The unperturbed eigenstate corresponding to
the $E^0_{300}$ displays localization properties similar to the
state shown in Fig.~\ref{fig1}b. The insets show the
quantum-classical correspondence for the average.}
\vskip0.5cm
\end{figure}

Since most eigenstates display similar statistical features, an
average considering neighbouring eigenstates will smooth the
fluctuations, and we expect a better correspondence. In
Figs.~\ref{fig2} we present the results for the EF and the LDOS
obtained for a recentered average of eigenstates containing 21
eigenstates. The average was performed by recentering each
eigenstate, so the peak associated with the actual energy of the
eigenstate is labeled as zero. We notice that this procedure
incorporates the fact that neighbouring eigenstates are typically
similar.

As shown in the insets of Figs.~\ref{fig1}-\ref{fig2}, for 
generic eigenstates there is a clear correspondence 
between the classical and the quantum results. As it can be
appreciated in the plots, in this case, the tails of the
distribution are well approximated by the classical calculations,
while the central peak is the main concern for the correspondence.
The results presented are actually improved as the semiclassical
limit is reached. In fact, as we approach the semiclassical
limit, the density of states increases and therefore the central
classical peak is better resolved. As one would expect, the
localized eigenstates display strong deviations from the classical
results. Notice though, that after averaging the classical
correspondence emerges again, since the main contributions come
from the (typical) extended eigenstates. This is appreciated in
Fig.~\ref{fig2}b, where we present the LDOS obtained for the
recentered average taken over 21 eigenstates, where we have chosen
the central eigenstate to be localized.

At this point we shall emphasize that no free parameter has been
used to fit the data. Namely, the classical energy is taken from
the energy of the eigenstate under consideration (and the scaling
property is used); for the results involving the average, the
energy corresponds to the average energy of the eigenstates within
the window. Furthermore, the similarity (under certain reflection)
of the LDOS and the EF displayed in Figs.~\ref{fig2} is a
consequence of the symmetry of $H$ and $H_0$ in
Eq.~(\ref{classical}), as given by~(\ref{ldos}).

The good correspondence found for the EF and the LDOS in the
semiclassical limit can be understood by interpreting
Eq.~(\ref{classical}) as a kind of generalization of the Weyl
formula for the intensities of the eigenstates, with respect to a
certain basis $H_0$. The fact that this expression is a phase
space integral, implies that it contains no information about the
integrability or chaos of the classical systems.

While the above quantum-classical correspondence for the LDOS and EF,
to a large extent, can be expected from previous studies of other 
models~\cite{izr2}, in what follows we concentrate on the analysis of
``single-particle'' properties, in particular, the occupation
numbers of single-particle states (see also \cite{fausto} where
this quantity was studied for two interacting spins). This may
seem of marginal importance for two-particle systems, though it is
certainly of great significance as the number of particles
increases \cite{fausto2}. Yet, the two-particle system will be an
adequate test ground to study how the 
semiclassical approach can be applied.

In terms of the expansion coefficients $C^i_k$, the single-particle
occupation number distribution $n_s$ is defined as
\beq
\label{ns}
<n^i_s>\equiv \langle \Psi_i|\hat{n}_s|\Psi_i\rangle=
\sum_k |C_k^i|^2 n_s^{(k)},
\eeq
where $\hat{n}_s=a^\dagger_s a_s$ is the occupation number operator,
$a_s^\dagger$ and $a_s$ are the creation and annihilation operators,
and $n_s^{(k)}=\langle\Phi^0_k|a_s^\dagger a_s|\Phi^0_k\rangle$.
Aside from its significance in statistical mechanics, 
the interest of the occupation number operator 
is that it allows to calculate
mean values of any single-particle operator $<M>=\sum_s n_s M_{ss}$.

The classical $n_s$ is defined in the same terms of the classical EF
or LDOS.
Accordingly, it is obtained by computing the time dependent
single-particle energy distribution $\epsilon(t)$, using the solutions
of the classical equations of motion for $H$.
Again, an expression similar to Eq.~(\ref{classical}) can be written
for the classical occupation number distribution, which is given by
\beq
\label{classicalns}
g_n({\epsilon}, {\cal E})=
A^\prime \int {\rm d}{\bf p}{\rm d}{\bf q}
\delta({\cal E}-H({\bf p},{\bf q}))
\sum_{i=1}^2
\delta({\epsilon}-h_i(p_i,q_i)).
\eeq
Here, $\epsilon$ is the independent variable, ${\cal E}$ is the
energy of the full Hamiltonian, and $h_i$ represents the one-particle
Hamiltonian.
In the present case we have assumed the particle interchange symmetry,
so the sum in Eq.~(\ref{classicalns}) may be absorbed in the
normalization constant, which corresponds to the number of particles.

\begin{figure}
\noindent\centerline{
\psfig{figure=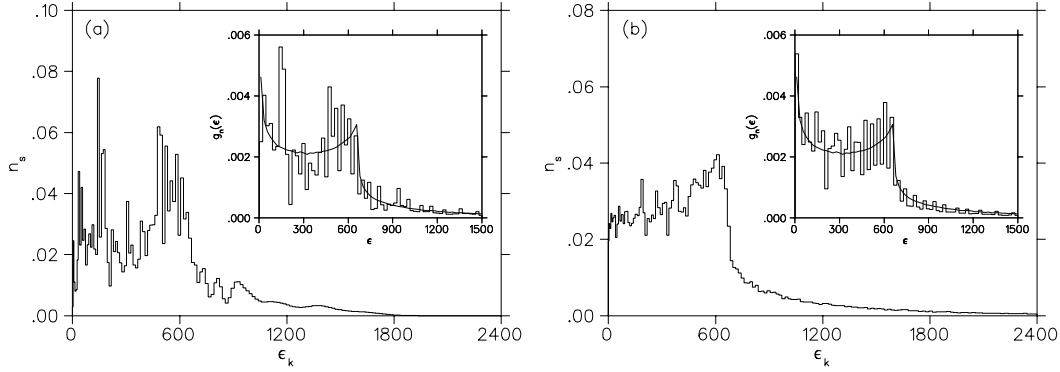,height=\textwidth,angle=90}}
\caption{
\label{fig3}
Quantum-classical correspondence for the distribution of occupation 
numbers $n_s$. The eigenstate considered here is the one shown in 
Fig.~\ref{fig1}a. The calculations for bosons display a good agreement 
too, and also an interesting similarity to those presented here 
for fermions.}
\vskip0.5cm
\end{figure}

In Fig.~\ref{fig3} we present the results for $n_s$. No free parameter 
was used to fit the data. A good correspondence of the classical and 
quantum results is found, although the calculations that involve 
individual eigenfunctions display very large fluctuations. The 
correspondence is certainly better if we perform an average over some
neighbouring eigenstates, and again is improved as we go deeper
into the semiclassical region. It is not clear, however, to what
extent the fluctuations observed for individual eigenfunctions are
related to quantum localization effects.

The correspondence shown in Figs.~\ref{fig3} is novel and important. 
First, it has no especial interpretation in the framework of 
classical mechanics for isolated systems of few interacting 
particles, although the single-particle occupation number
distribution is an important quantity in quantum statistical
mechanics, where is linked to the Boltzmann distribution (in the
thermodynamic limit). Second, we note that the tail of the 
$n_s$-distribution, which displays exponential decay, is well 
reproduced by Eq.~(\ref{classicalns}). This is a non-trivial remark 
if we recall that we deal with a two-particle system. Clearly, 
this permits to define an analogous of the Boltzmann parameter for 
finite systems, although its interpretation as the inverse of the 
temperature is not generically accepted (see the discussion 
in~\cite{fausto2}).

Once we have shown that good quantum-classical correspondence is found 
in the semiclassical limit, it is possible to proceed with estimates
of other quantities involving the F-function. Our calculations are
based on the classical EF as calculated above, and require only the 
knowledge of the single-particle spectra (that allows to compute the 
unperturbed spectra) and the perturbed spectra. We shall refer our 
prescription as ``semiquantum approach''. Specifically, we illustrate 
the method by a calculation of the inverse participation ratio (IPR),
\beq
\label{pratio}
P^+(E_i)= \sum_k |C_k^i|^4,
\eeq
which is an important measure of the uniformity of the expansion
distribution~\cite{BSW93,mejiaetal}. Other quantities which involve 
even powers of the expansion coefficients, i.e. where the phase of
eigenfunctions does not appear, can be obtained in the same terms.

In order to compute the IPR, we must express our classical
(continuous) distribution as an intensity distribution, which is
normalized as an eigenfunction. Obviously, our results will depend
on how we discretize this distribution, though no free parameter
will be involved. The comparison between the semiquantum and the
quantum results will thus give insight in the plausibility of the
discretization. One possibility for this discretization is the
following: We divide the unperturbed energy range into segments,
such that every segment contains only one unperturbed eigenvalue,
and that they span the whole interval. We define the limits of
each such interval by the middle point of neighbouring eigenvalues. 
Then, we define the
classical intensity associated with an unperturbed energy as the
value of the area under the classical distribution corresponding
to the segment, that contains the unperturbed eigenvalue. This
procedure leads to a semiquantum intensity distribution with the
required normalization, and can, therefore, be used to obtain
quantities like the $n_s$ or a semiquantum version of the IPR.
However, the semiquantum intensities obtained in this way will
display more zeros than the corresponding quantum ones. This is an
obvious consequence of the finite range where the classical
density $g({\cal E},{\cal E}_0)$ is non-zero.

\begin{figure}
\noindent\centerline{
\psfig{figure=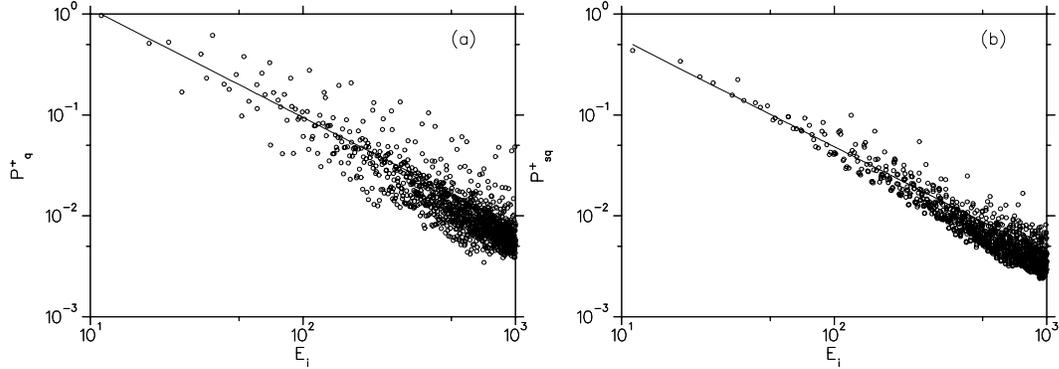,height=\textwidth,angle=90}}
\caption{
\label{fig4}
IPR as a function of the energy calculated for individual eigenstates;
(a)~quantum and (b)~semiquantum calculations. The straight lines 
plotted correspond to the best fit of a power-law. The differences on 
the slopes of the semiquantum and quantum results are less that 1\%.}
\vskip0.5cm
\end{figure}

We note that from the classical scaling properties of the EF,
$g({\cal E},{\cal E}_0)\sim {\cal E}^{-1}$, and the mean level 
density, $\rho({\cal E})\sim {\cal E}^{1/2}$, we can obtain a 
semiquantum estimate for the IPR, which is expected to display a 
power-law decay of the form
$[g({\cal E},{\cal E}_0) \rho({\cal E})]^2\sim {\cal E}^{-1}$. In
Figs.~\ref{fig4} we present results for the semiquantum and
quantum IPR. The quantum results shown correspond to the direct
evaluation of Eq.~(\ref{pratio}) for individual eigenstates, that
is, no average or smoothing prescription has been used. A
qualitative agreement is observed between the semiquantum and
the quantum data, although quantitative differences arise. The
power-law decay predicted from the classical scaling properties of
the system~(\ref{model}) is well confirmed. In fact, we have
fitted curves of the form $P^{+}_i \propto E_i^{-\mu}$ to our
results, and obtained that the best fit is provided by
$\mu_{sq}=1.06$ and $\mu_q=1.07$ for the semiquantum and quantum
data, respectively. Analogue results, obtained when considering
averages over windows of the eigenstates, show better quantitative
agreement. In this case, the features related with large values of
the IPR (localized eigenstates) are smeared out.

It is interesting to note that certain semiquantum states display
a rather large IPR, which could be associated with some
localization properties (scars). In turn, the quantum results
display more of these localized states and the IPR associated is
also larger. This enhancement of localization is a well-known
quantum effect.

In summary, we have shown that important quantities like the EF,
LDOS or $n_s$ have, in the semiclassical limit, good correspondence 
to their classical analogues in our model of two interacting 
non-linear oscillators. The classical quantities are obtained as 
phase space integrals, assuming that they are applied to the case 
when the classical dynamics is strongly chaotic. We have used the 
classical quantities, in particular the classical EF, together with 
the quantum spectra of the perturbed and unperturbed systems in 
order to obtain ``semiquantum'' intensities associated to a given
(perturbed) eigenstate. We have compared the quantum results with
the semiquantum ones, specifically for the inverse participation
ratio, and found good qualitative agreement, although
quantitatively they may display differences. In particular, the
semiquantum results underestimate the IPR for rather localized
(scars) eigenstates, both in their number and magnitude. The
energy dependence of the IPR can be understood from the classical
scaling properties of our model.

Since quantities like the IPR involve information on the underlying 
classical mechanics, we believe that semiquantum properties may help 
to explore quantum localization effects in the semiclassical region. 
Moreover, expressions like Eq.~(\ref{classical}) may help to define 
a reference from which the study of eigenfunction fluctuations 
and their relation to the underlying dynamics can be achieved. 
Our results directly take into account the two-body nature of 
the inter-particle interaction, and can be easily extended to 
systems with any number of particles.

\ack
The authors want to express sincerely their gratitude for the
useful discussions to Fran\c{c}ois Leyvraz. This work was
partially supported by the DGAPA (UNAM) project IN-102597 and the
CONACYT (M\'exico) Grants No. 25192-E, 26163-E and No. 28626-E.


\end{document}